\documentclass[12pt,preprint]{aastex}               \usepackage{natbib}
\usepackage{bm}

\def\be{\begin{equation}}
\def\ee{\end{equation}}
\begin{document}
\title{Fast Estimator of Primordial Non-Gaussianity from Temperature and
Polarization Anisotropies in the Cosmic Microwave Background II: Partial
Sky Coverage and Inhomogeneous Noise}

\author{     Amit    P.    S.     Yadav\altaffilmark{1,2},    Eiichiro
Komatsu\altaffilmark{3},   Benjamin   D.   Wandelt\altaffilmark{2,4,5},
Michele  Liguori\altaffilmark{6}, \\  Frode  K.  Hansen\altaffilmark{7},
Sabino        Matarrese\altaffilmark{8}}       \altaffiltext{1}{email:
ayadav@uiuc.edu} \altaffiltext{2} {Department of Astronomy, University
of  Illinois at  Urbana-Champaign,  1002 W.~Green  Street, Urbana,  IL
61801, USA}  \altaffiltext{3}{ Department of  Astronomy, University of
Texas at Austin, 2511 Speedway, RLM 15.306, TX 78712, USA}

\altaffiltext{4}{  Department of  Physics, University  of  Illinois at
Urbana-Champaign, 1110 W.~Green Street, Urbana, IL 61801, USA}

\altaffiltext{5}{Center of Advanced  Studies, University of Illinois at
Urbana-Champaign, 912 W.~Illinois Street, Urbana, IL 61801, USA}

\altaffiltext{6}{Department  of  Applied  Mathematics and  Theoretical
Physics,   Center    for   Mathematical   Sciences,    university   of
Cambridge,Wilberfoce  Road,   Cambridge,  CB3  0WA,   United  Kingdom}
\altaffiltext{7}{Institute of  Theoretical Astrophysics, University of
Oslo,    P.O.   Box    1029    Blindern,   0315    Oslo,   Norway    }
\altaffiltext{8}{Departmento di Fisica  ``G. Galilei'', Universit\`a di
Padova and  INFN, Sezione di  Padova, via Marzolo 8,  I-35131, Padova,
Italy}

\begin{abstract}
In our recent paper~\citep{YKW07} we described a fast cubic (bispectrum)
 estimator of the amplitude of primordial non-Gaussianity of local type,
 $f_{NL}$, from a combined analysis of the Cosmic Microwave Background
 (CMB) temperature and E-polarization observations.  
 In this paper we generalize the estimator to deal with a partial sky
 coverage as well as inhomogeneous noise. Our generalized estimator 
is still computationally efficient, scaling as $O(N^{3/2}_{pix})$ compared to
the $O(N^{5/2}_{pix})$ scaling of the brute force bispectrum calculation for
sky maps with $N_{pix}$ pixels.  
Upcoming CMB experiments are expected to yield high-sensitivity
 temperature and E-polarization data. Our
 generalized estimator will allow us to optimally utilize the combined
 CMB temperature and E-polarization information from these realistic
 experiments, and to constrain primordial non-Gaussianity. 
\end{abstract} 

\section{Introduction} 
\label{introduction}

Non-Gaussianity from the simplest inflation models, that are based on a
slowly        rolling        scalar        field,        is        very
small~\citep{Salopek_Bond90,Salopek_Bond91,Falk_et93,Gangui_et94,Acquaviva02,Maldacena03};
however,  a  very large  class  of  more  general models, e.g., models with
multiple scalar fields, features in inflation potential, non-adiabatic
fluctuations,    non-canonical   kinetic   terms,    deviations   from
the Bunch-Davies vacuum, among others, predict  substantially   higher
level of primordial non-Gaussianity \cite[for a review
and references therein]{BKMR_04}.

Primordial non-Gaussianity can be described in terms of the 3-point
correlation function of Bardeen's curvature perturbations, $\Phi(k)$, in
Fourier space: 
\begin{eqnarray}
\langle \Phi(\mathbf{k_1})(\mathbf{k_2})(\mathbf{k_3})\rangle = (2\pi)^3\delta^3(\mathbf{k_1} + \mathbf{k_2} + \mathbf{k_3})F(k_1, k_2, k_3).
\end{eqnarray}
Depending on the shape of the 3-point function, i.e., $F(k_1, k_2,
k_3)$, non-Gaussianity can be broadly classified into two
classes~\citep{Babich_etal_04}. 
First, the local, ``squeezed,'' non-Gaussianity where
$F(k_1, k_2, k_3)$ is large for the configurations in which $k_1 \ll k_2,
k_3$. Second, the non-local, ``equilateral,'' non-Gaussianity where $F(k_1, k_2, k_3)$ is
large for the configuration when $k_1 \sim k_2 \sim k_3$. 

The local form arises from a non-linear relation between
inflaton and curvature
perturbations~\citep{Salopek_Bond90,Salopek_Bond91,Gangui_et94},
curvaton models~\citep{Lyth_Ungarelli_Wands_03}, 
or the New Ekpyrotic models~\citep{Koyama_etal07,Buchbinder_etal07}. 
The equilateral form arises from non-canonical kinetic terms such 
as the Dirac-Born-Infeld (DBI)
action~\citep{Alishahiha_etal04}, the ghost
condensation~\citep{Arkani_et_04}, or any other single-field models in
which the scalar field acquires a low speed of sound
~\citep{Chen_etal07,Cheung_Creminelli_etal07}.
While we focus on the local form in this paper, it is straightforward to
repeat our analysis for the equilateral form.

The local form of non-Gaussianity may be parametrized in real space
as~\citep{Gangui_et94,verde00,KS2001}: 
\begin{equation}
\label{eqn:phiNG}
\Phi(\mathbf{r}) = \Phi_L(\mathbf{r}) + f_{NL} \left( \Phi_L^2(\mathbf{r}) - \langle \Phi_L^2(\mathbf{r}) \rangle
\right)
\end{equation}
where $f_{NL}$ characterizes the amplitude of primordial
non-Gaussianity. 
Different
inflationary  models predict  different amounts  of  $f_{NL}$, starting
from $O(1)$ to $f_{NL}\sim 100$, beyond which values have been excluded
by the Cosmic Microwave Background (CMB) bispectrum of WMAP temperature
data, $-36 < f_{NL} <100$, at the $2\sigma$ level
\citep{nong_wmap,creminelli_wmap2,wmap_2nd_spergel}.

So far all the
constraints on primordial non-Gaussianity use 
only temperature information of the CMB. 
By also  having the E-polarization  information together with CMB temperature information, one  can  improve the sensitivity to the primordial  fluctuations
\citep{BZ04,YW05,YKW07}. Although the  experiments have already started
characterizing    E-polarization    anisotropies    \citep{dasi_pol_02,
wmap_1st_pol, wmap_2nd_pol,boom_ee},  the   errors   are  large   in
comparison to temperature anisotropy. The upcoming experiments such as
Planck satellite will characterize E-polarization anisotropy to a higher
accuracy. It is very timely to develop the tools which can optimally
utilize the combined CMB temperature and E-polarization information to
constrain models of the early universe.    

Throughout this paper we use the standard Lambda CDM cosmology with the
following cosmological parameters:  
$\Omega_b = 0.042$, $\Omega_{cdm} = 0.239$, $\Omega_L = 0.719$, 
$h = 0.73$, $n = 1$, and $\tau = 0.09$. For all of our simulations we
used HEALPix maps with $N_{pix}\approx 3\times 10^6$ pixels. 

\subsection{Generalized Bispectrum Estimator of Primordial Non-Gaussianity}
\label{estimator_generalized}
In our recent paper~\citep{YKW07} we  described  a   fast  cubic
(bispectrum)  estimator  of $f_{NL}$,  using  a  combined analysis  of the
temperature and E-polarization  observations.  The estimator was optimal
for homogeneous noise, where optimality was defined  by saturation of
the Fisher matrix bound. 

In this paper we generalize our previous estimator of $f_{NL}$
to deal more optimally with a partial sky coverage and the inhomogeneous
noise. The generalization is done in an analogous way to
how~\cite{Creminelli_wmap1} generalized the temperature only
estimator developed by \citet{KSW05}; however, the final result of
\cite{Creminelli_wmap1} (their Eq.~(30)) is off by a factor of two,
which results in the error in $f_{NL}$ that is much larger than the
Fisher matrix prediction, as we shall show below.

The fast bispectrum estimator of $f_{NL}$ from the combined CMB
temperature and E-polarization data can be written as 
$\hat{f}_{NL} = \frac{\hat{S}_{prim}}{N}$,
where \citep{YKW07}
\begin{eqnarray}
\label{s_prim}
\hat{S}_{prim}=\frac{1}{f_{sky}}\int     r^2dr     \int     d^2\hat{n}\;
B^2(\hat{n},r) A (\hat{n},r),
\end{eqnarray} 

\begin{eqnarray}
N=  \sum_{i j  k  p q
r}\sum_{2 \le \ell_1\le  \ell_2\le  \ell_3} \frac{1}{\Delta_{\ell_1 \ell_2 \ell_3}} B^{p q  r,prim}_{\ell_1  \ell_2
\ell_3}(C^{-1})^{ip}_{\ell_1}(C^{-1})^{j         q}_{\ell_2}(C^{-1})^{k
r}_{\ell_3} B^{i j k,prim}_{\ell_1 \ell_2 \ell_3 },
\end{eqnarray}

\begin{eqnarray}
\label{B}
B(\hat{n},r)\equiv           \sum_{ip}\sum_{lm}(C^{-1})^{ip}a^{i}_{\ell
m}\beta^p_{\ell}(r)Y_{\ell m}(\hat{n}),
\end{eqnarray}
\begin{eqnarray}
\label{A}
A(\hat{n},r)\equiv           \sum_{ip}\sum_{lm}(C^{-1})^{ip}a^{i}_{\ell
m}\alpha^p_{\ell}(r)Y_{\ell m}(\hat{n}),
\end{eqnarray}

\begin{eqnarray}\beta^i_{\ell}(r)=\frac  {2b^i_{\ell}}{\pi} \int k^2  dk\; P_{\phi}(k)g^{i}_{\ell}(k)\,
j_{\ell}(kr),
\end{eqnarray}

\begin{eqnarray}\alpha^i_{\ell}(r)=\frac  {2b^i_{\ell}}{\pi} \int k^2  dk\; g^{i}_{\ell}(k)\,
j_{\ell}(kr),
\end{eqnarray}
and $f_{sky}$ is a  fraction of the sky observed. Indices
$i, j, k, p, q$ and $r$ 
can  either be  $T$ or  $E$. Here, $\Delta_{\ell_1  \ell_2 \ell_3}$ is 1
when  $\ell_1 \neq \ell_2 
\neq \ell_3$, 6 when $\ell_1 =  \ell_2 = \ell_3$, and 2 otherwise, $B^{p q r,prim}_{\ell_1 \ell_2  \ell_3}$ is the
 theoretical bispectrum for $f_{NL}=1$~\citep{YKW07}, $P_{\Phi}(k)$
 is  the  power  spectrum of  the  primordial  curvature 
perturbations, and $g^i_{\ell}(r)$ is the radiation 
transfer function of adiabatic perturbations. 

It has been shown that the above mentioned estimator is optimal for the
full sky coverage and homogeneous noise \citep{YKW07}. To be able to
deal with the realistic data, the estimator has to be able to deal with
the inhomogeneous noise and foreground masks. 

The estimator can be
generalized to deal with a partial sky coverage and the inhomogeneous noise by
adding a linear term to $\hat{S}_{prim}$: $\hat{S}_{prim}\rightarrow 
\hat{S}_{prim}+\hat{S}_{prim}^{linear}$. 
For the temperature only case, this
has been done in~\cite{Creminelli_wmap1}. Following the same argument,
we find that 
the linear term for the combined analysis of CMB temperature and
polarization data is given by 
\begin{eqnarray}
\label{slin}
\hat{S}^{linear}_{prim}=\frac{-1}{f_{sky}}\int  r^2dr  \int d^2\hat{n}
\left   \{  2 B(\hat{n},r)\, \langle A_{sim}(\hat{n},r)B_{sim}(\hat{n},r)\rangle_{MC}  +     A(\hat{n},r)\, \langle B^2_{sim}(\hat{n},r) \rangle_{MC}
 \right \},
\end{eqnarray} 
where  $A_{sim}(\hat{n},r)$ and $B_{sim}(\hat{n},r)$ are the $A$ and $B$
maps generated from Monte Carlo simulations that contain signal and
noise,  and $\langle ..\rangle$ denotes the average over the Monte Carlo
simulations.

The generalized estimator is given by 
\begin{equation}
 \hat{f}_{NL} =
\frac{\hat{S}_{prim}+\hat{S}^{linear}_{prim}}{N},
\end{equation}
which is the main result of this paper.
Note that
$\langle\hat{S}^{linear}_{prim}\rangle_{MC}=-\langle\hat{S}_{prim}\rangle_{MC}$,
and this relation also holds for the equilateral shape. Therefore, it is
straightforward 
to find the generalized estimator for the equilateral shape: first, find
the cubic estimator of the equilateral shape, $\hat{S}_{equilateral}$,
and take the Monte Carlo average,
$\langle\hat{S}_{equilateral}\rangle_{MC}$. 
Let us suppose that $\hat{S}_{equilateral}$ contains terms in the form
of $ABC$, where $A$, $B$, and $C$ are some filtered maps.
Use the Wick's theorem to
re-write the average of a cubic product as 
$\langle ABC\rangle_{MC}=
\langle A\rangle_{MC}\langle BC\rangle_{MC}
+\langle B\rangle_{MC}\langle AC\rangle_{MC}
+\langle C\rangle_{MC}\langle AB\rangle_{MC}$. Finally, 
remove the MC average from single maps, and replace maps in the product
with the simulated maps: 
$\langle A\rangle_{MC}\langle BC\rangle_{MC}
+\langle B\rangle_{MC}\langle AC\rangle_{MC}
+\langle C\rangle_{MC}\langle AB\rangle_{MC}
\rightarrow
A\langle B_{sim}C_{sim}\rangle_{MC}
+B\langle A_{sim}C_{sim}\rangle_{MC}
+C\langle A_{sim}B_{sim}\rangle_{MC}$. This operation gives the correct
expression for the linear term, both for the local form and the
equilateral form.

One can find the estimator of $f_{NL}$ from the temperature data
 only by setting $i=j=k=p=q=r\equiv T$.
We have compared our formula in the temperature-only limit with  the
 original formula derived by~\citet{Creminelli_wmap1} (their Eq.~(30)),
 and found a discrepancy. To see the discrepancy, let us re-write the
 estimator as: $\hat{f}_{NL} = 
\frac{\hat{S}_{prim}+ x \hat{S}^{linear}_{prim}}{N}$. Our formula gives 
$x=1$, while Eq.~(30) of \citet{Creminelli_wmap1} gives $x=0.5$.
\footnote[1]{Equation~(30)
in~\citet{Creminelli_wmap1} is off by a factor  of 2. Since we used  $\sum_{\ell_1 \ell_2 \ell_3}  = 6\sum_{\ell_1 \le
\ell_2  \le  \ell_3} \frac{1}{\Delta_{\ell_1 \ell_2 \ell_3}}$, to
compare our normalization factor $x$ with~\citet{Creminelli_wmap1} one
needs to divide the normalization in~\citet{Creminelli_wmap1} by a factor
of $6$.}  

\begin{figure*}[t]
\begin{center}
\includegraphics[height=8cm,angle=0]{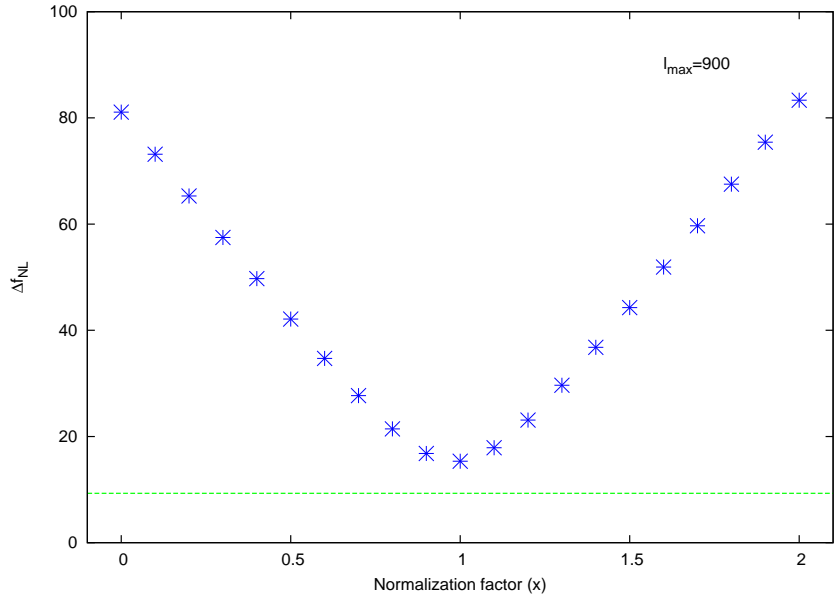}
\caption{Testing normalization of the linear term in the estimator of
 $f_{NL}$. The symbols show the standard deviation of $f_{NL}$ derived from the
 the Monte Carlo simulations using the estimator for a given
 normalization, $x$. The horizontal line shows 
 the Fisher matrix prediction. Our formula gives $x=1$, while  
\citet{Creminelli_wmap1} give $x=0.5$ (their Eq.~(30)). 
 We have used simulated polarized Gaussian CMB maps with the Planck
 inhomogeneous 
 noise as well as the WMAP Kp0 and P06 mask for temperature and
 polarization, respectively.} 
\label{normalization}
\end{center}
\end{figure*}

To make sure that our normalization gives the minimum variance
estimator, we have done Monte Carlo simulations
with varying $x$. We find that $x=1$ minimizes
the variance (as shown in Fig.~\ref{normalization}). 
We conclude that the analysis given in \citet{Creminelli_wmap1} resulted
in the larger-than-expected uncertainty in $f_{NL}$ because of this
error in their normalization of the linear term.

\begin{figure*}[t]
\begin{center}
\includegraphics[height=5.5cm,angle=0]{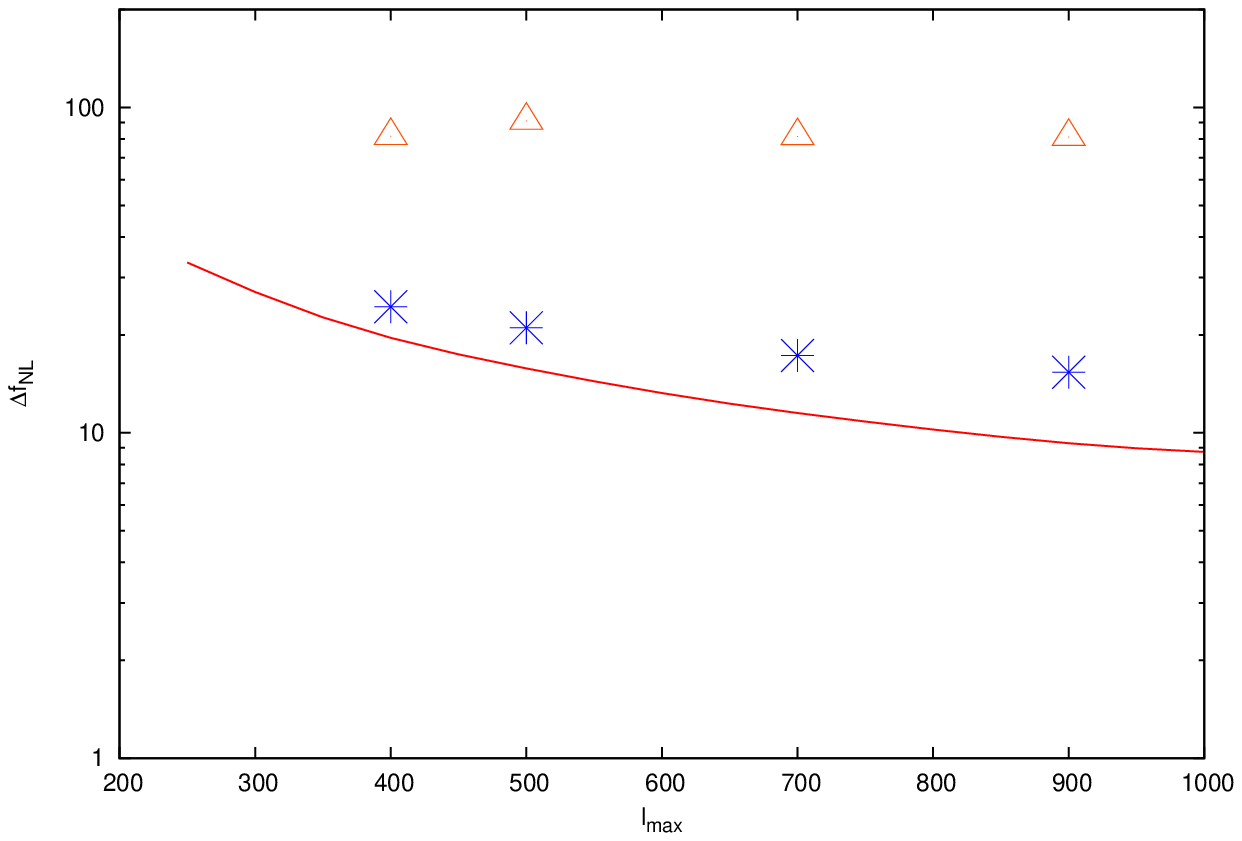}
\includegraphics[height=5.5cm,angle=0]{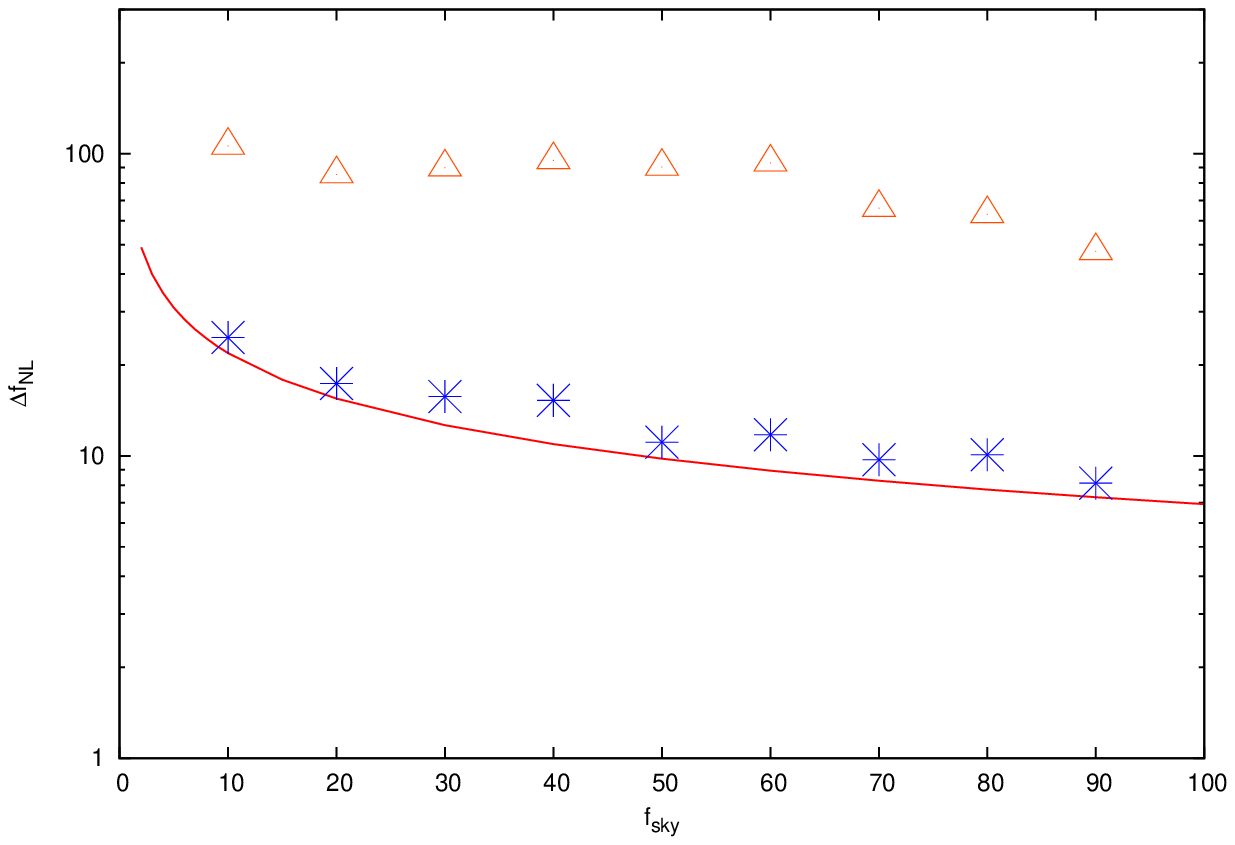}
\caption{Optimality of the generalized estimator.
The solid  lines  show  the  Fisher matrix prediction for the standard
 deviation of $f_{NL}$, the triangles
show the standard   deviation derived from the  Monte    Carlo
simulations using the estimator without the linear term, and the stars
 show the  standard   deviation  derived from the  Monte    Carlo
simulations using the generalized estimator (i.e., with the linear
 term). {\bf Left panel:} The uncertainty vs the maximum multipole that
 is used in the analysis, $\ell_{max}$. The simulations contain the
 Gaussian CMB signal, inhomogeneous noise (which simulates the Planck
 satellite), WMAP Kp0 and P06 masks. {\bf Right panel:} The uncertainty
 vs a fraction of the sky observed, $f_{sky}$, for $\ell_{max}=500$. The
 simulations include 
 the Gaussian CMB signal, and flat sky-cut (which is azimuthally symmetric
 in the Galactic coordinates), while they do not include instrumental
 noise. This  figure therefore shows that the sky cut contributes
 significantly to the linear term of polarization. 
}
\label{fnl_stdev}
\end{center}
\end{figure*}

 The main  contribution to the
linear term  comes from  the inhomogeneous noise and sky cut.
For the
temperature only 
case, most of the contribution to the linear term comes from the
inhomogeneous noise, and the partial sky coverage does not contribute
much to the linear term. This is because the sky-cut 
induces a monopole contribution outside the mask. In the analysis one subtracts the monopole
from outside the mask before measuring $\hat{S}_{prim}$, which makes
the linear contribution from the mask small \citep{Creminelli_wmap1}. 
For a combined analysis of the temperature and polarization maps,
however, the linear term does  get a
significant contribution from a partial sky coverage (see
the right panel of Fig.~\ref{fnl_stdev}). Subtraction of the monopole outside of the mask
is of no help for polarization, as the monopole does not exist in the
polarization maps by definition. (The lowest relevant multipole for
polarization is $l=2$.)

The estimator is still computationally efficient, taking only
$N^{3/2}_{pix}$ (times the $r$ sampling, which is of order 100) operations in comparison  to the full bispectrum calculation
which takes $N^{5/2}_{pix}$ operations. Here  $N_{pix}$ refers to the total number
of  pixels.  For Planck,  $N_{pix}  \sim  5 \times  10^7$,  and  so the  full
bispectrum analysis is not feasible while our analysis is.

\section{Results}

\begin{figure*}[t]
\begin{center}
\includegraphics[height=8cm,angle=90]{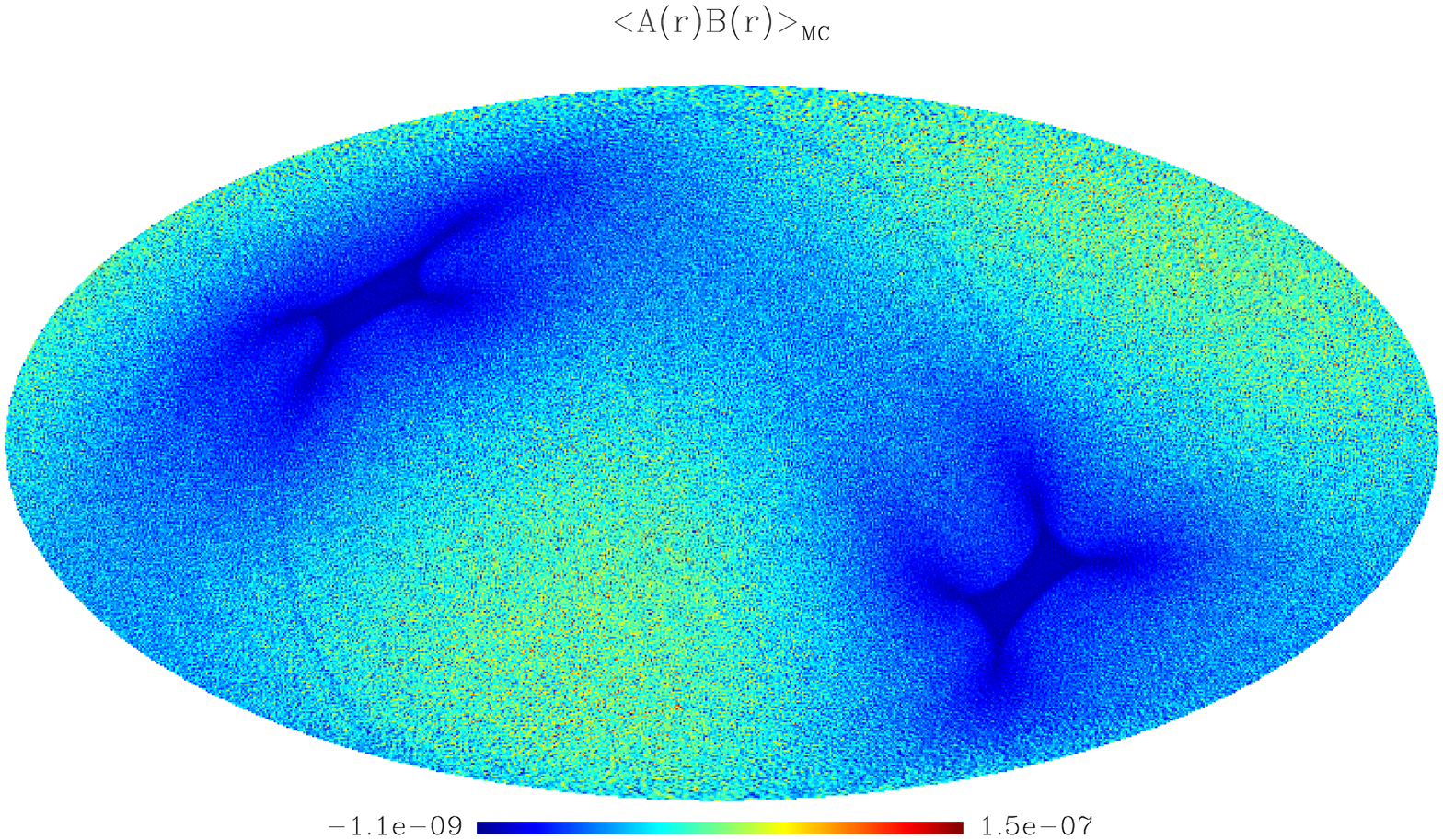}
\includegraphics[height=8cm,angle=90]{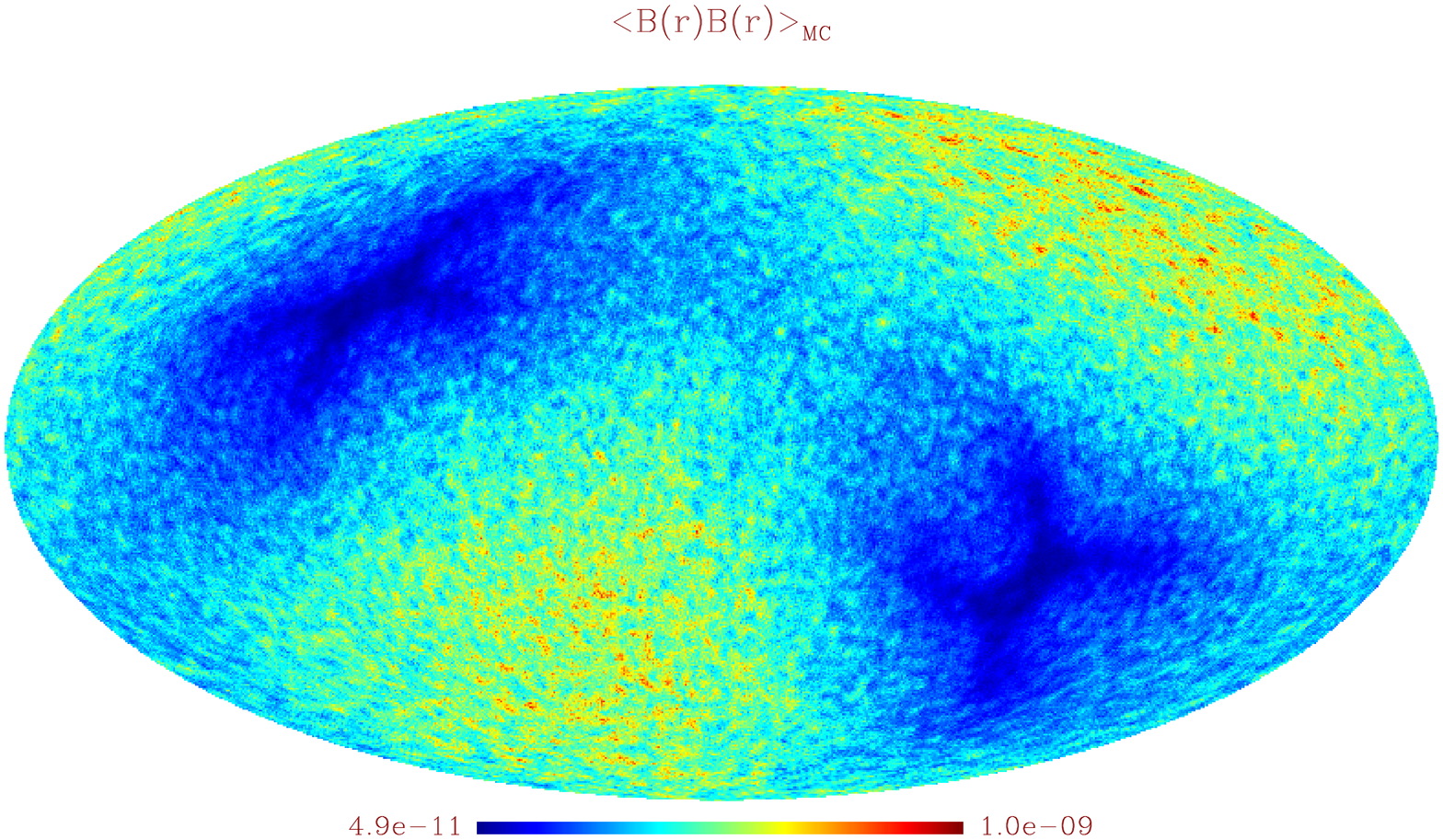}
\includegraphics[height=8cm,angle=90]{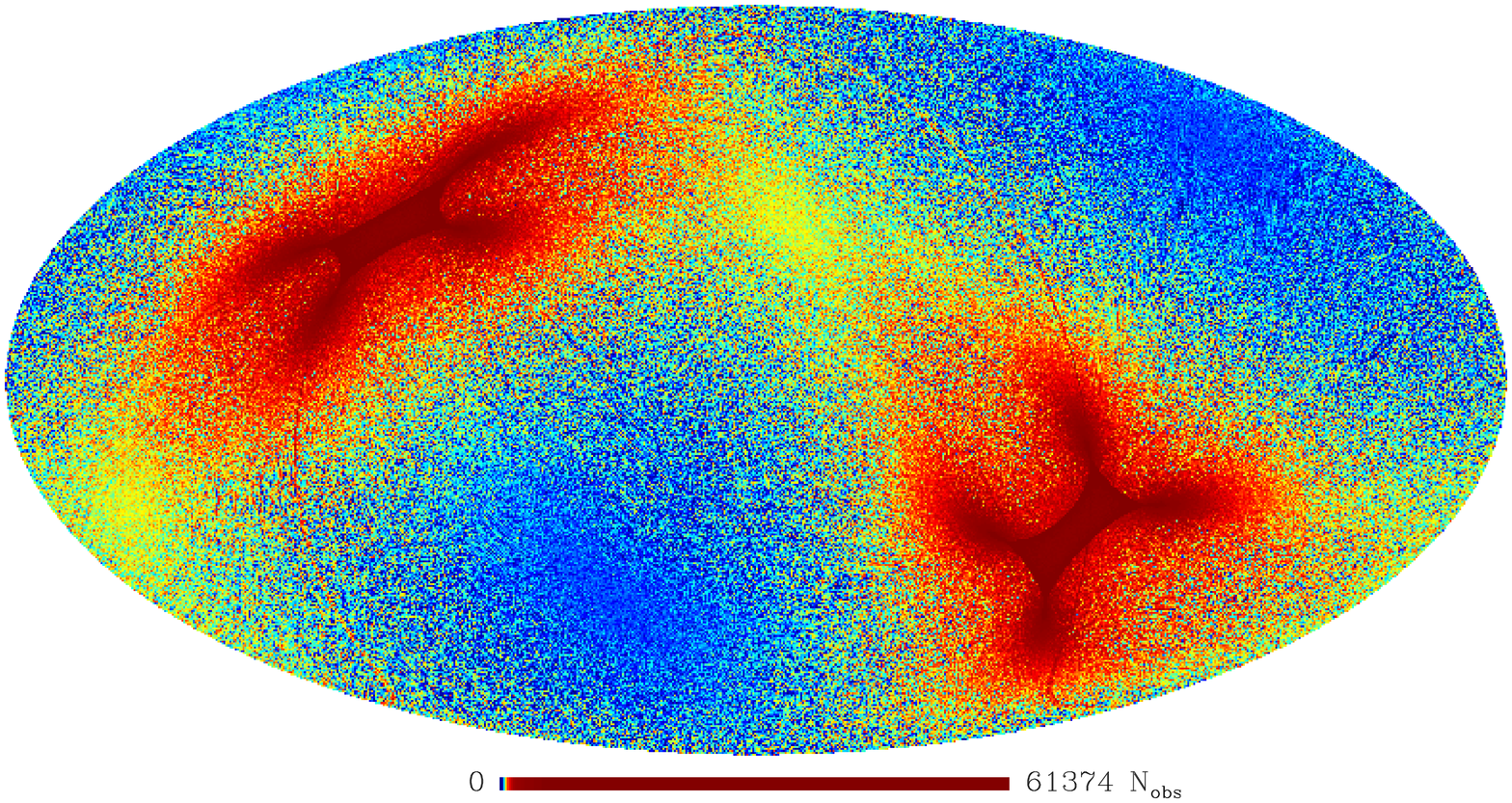}
\caption{The top panels show $\langle A_{sim}(\hat{n},r)B_{sim}(\hat{n},r)\rangle_{MC}$ and  $\langle B_{sim}^2(\hat{n},r)\rangle_{MC}$ maps for the noise only analysis (i.e. no CMB signal or mask). The maps are in dimension-less units and are shown for a slice near the  surface of  last
scattering. The Lower map shows the number of observations per pixel ($N_{obs}$) at the resolution of $N_{pix}=12582912$.}
\label{noise_sabb}
\end{center}
\end{figure*}

 In the left panel of figure~\ref{fnl_stdev}, we show the variance of
 $f_{NL}$ using the estimator (with and without the linear term) for the
 Gaussian CMB simulations in the presence of inhomogeneous noise and
 partial sky coverage. For this analysis we use the
 noise properties that are expected for the Planck satellite, assuming
 the cycloidal scanning strategy~\citep{Dupac_etal05}. Inhomogeneous
 nature of the noise is depicted in the lower map of
 figure~\ref{noise_sabb} where we show the number of observations
 ($N_{obs}$) for the different pixels in the sky. As for the foreground
 masks, we use WMAP Kp0 intensity mask and P06
 polarization mask. 
We find that, with the inclusion of the linear
 term, the variance reduces by more than a factor of 5. The linear term
 greatly reduces the variance approaching the Fisher matrix
 bound. However, the estimator is close to, but not exactly the same as
 the Fisher variance prediction in the noise dominated
 regime. 

Nevertheless, we do not observe the increase of variance at higher
$l_{max}$: the variance becomes smaller as we include more multipoles.
This result is in contradiction with the result of \citet{Creminelli_wmap1} and
\citet{creminelli_wmap2}. We attribute this discrepancy to the error in
the normalization of linear term in their formula. 

In the right panel of figure~\ref{fnl_stdev}, we show the variance of
$f_{NL}$ again using Gaussian simulations, but now in the presence of
flat sky cut and in the {\it absence} of any noise. The purpose of the plot is to demonstrate (as pointed out in the previous section) that for the combined CMB temperature and polarization analysis, sky-cut does contribute significantly to the linear term. We find that the generalized estimator does a very good job in reducing the variance excess, and the simulated variance of $f_{NL}$ does accurately saturate the Fisher matrix bound. 

Can our estimator recover the correct $f_{NL}$, i.e., is our estimator
unbiased? 
We have tested our estimator against simulated non-Gaussian CMB
temperature and E-polarization maps. 
The non-Gaussian CMB
temperature and E-polarization maps were generated using the method
described in~\cite{Liguori_Yadav_etal07}. 
We find that our estimator is
unbiased, i.e., we can recover the $f_{NL}$ value which was used to
generate the non-Gaussian CMB maps. The results for the unbiasedness of
the estimator are shown in Table~\ref{nonG_fnl}. The analysis also shows the
unbiasedness of the estimator described in~\cite{YKW07}.

\begin{table}[t]

\begin{center}
\begin{tabular}{|c|c|c|c|c|}
\hline
{\bf Noise} & {\bf Sky-cut} & $\mathbf{\langle f_{\rm NL}
  \rangle}$ & $\mathbf{f^{input}_{\rm NL}}$  & $\mathbf{\sigma_{sim}}$    \\
\hline
No & flat cut, $f_{sky}=0.8$ & $103.2$ & $100$ &$10.1$  \\
\hline
Inhomogeneous & WMAP Kp0 and P06 masks& $108.7$  & $100$ &$21.04$ \\
\hline
\end{tabular}\caption{Unbiasedness of the generalized
 estimator. Non-Gaussian CMB maps with $\mathbf{f^{input}_{\rm NL}}=100$
 are used for $\ell_{max}=500$. The standard deviation of $f_{NL}$,
 $\sigma_{sim}$, was obtained using Gaussian simulations.}
\label{nonG_fnl}
\end{center}
\end{table}

\begin{figure*}[t]
\begin{center}
\includegraphics[height=8cm,angle=90]{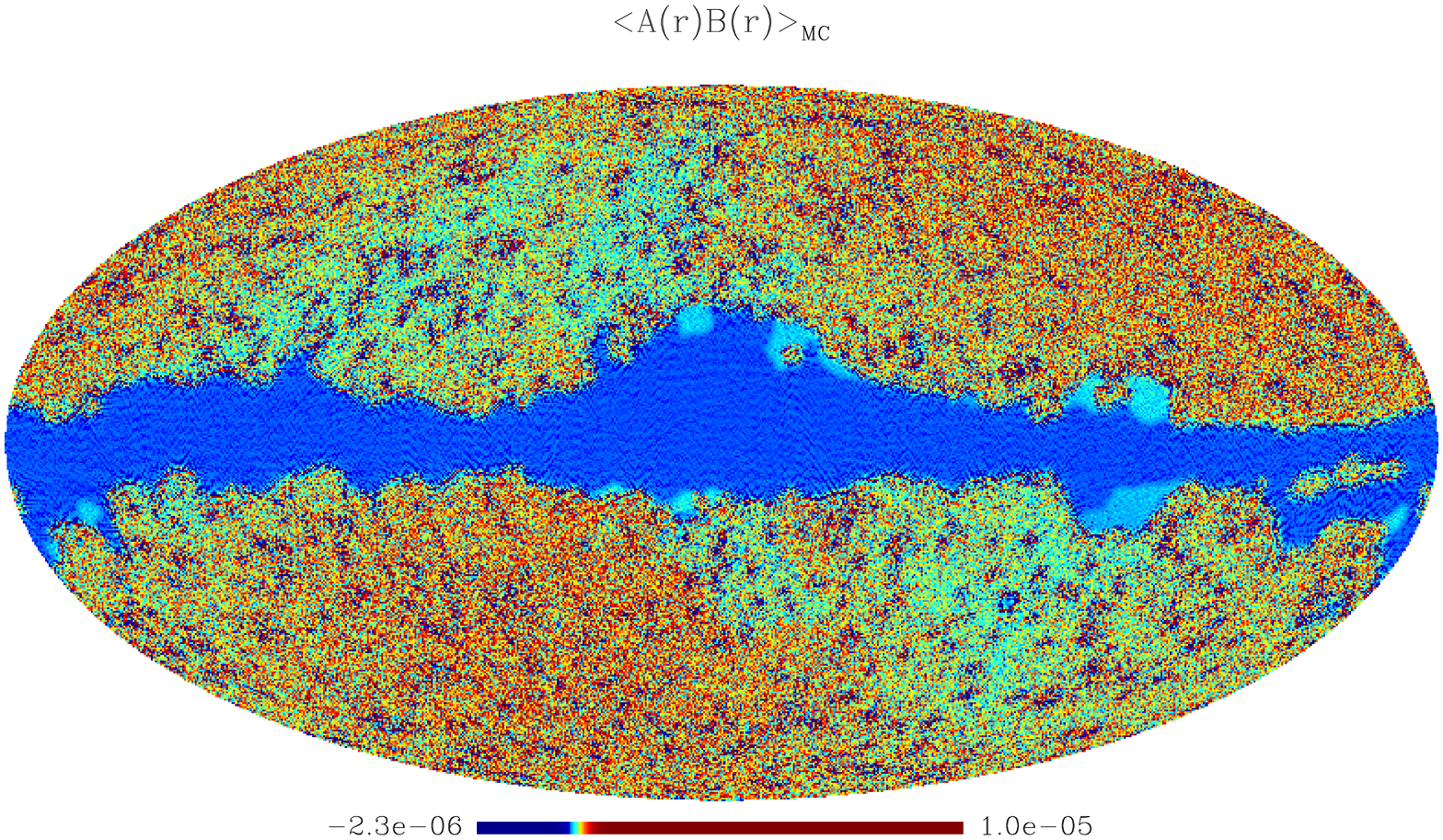}
\includegraphics[height=8cm,angle=90]{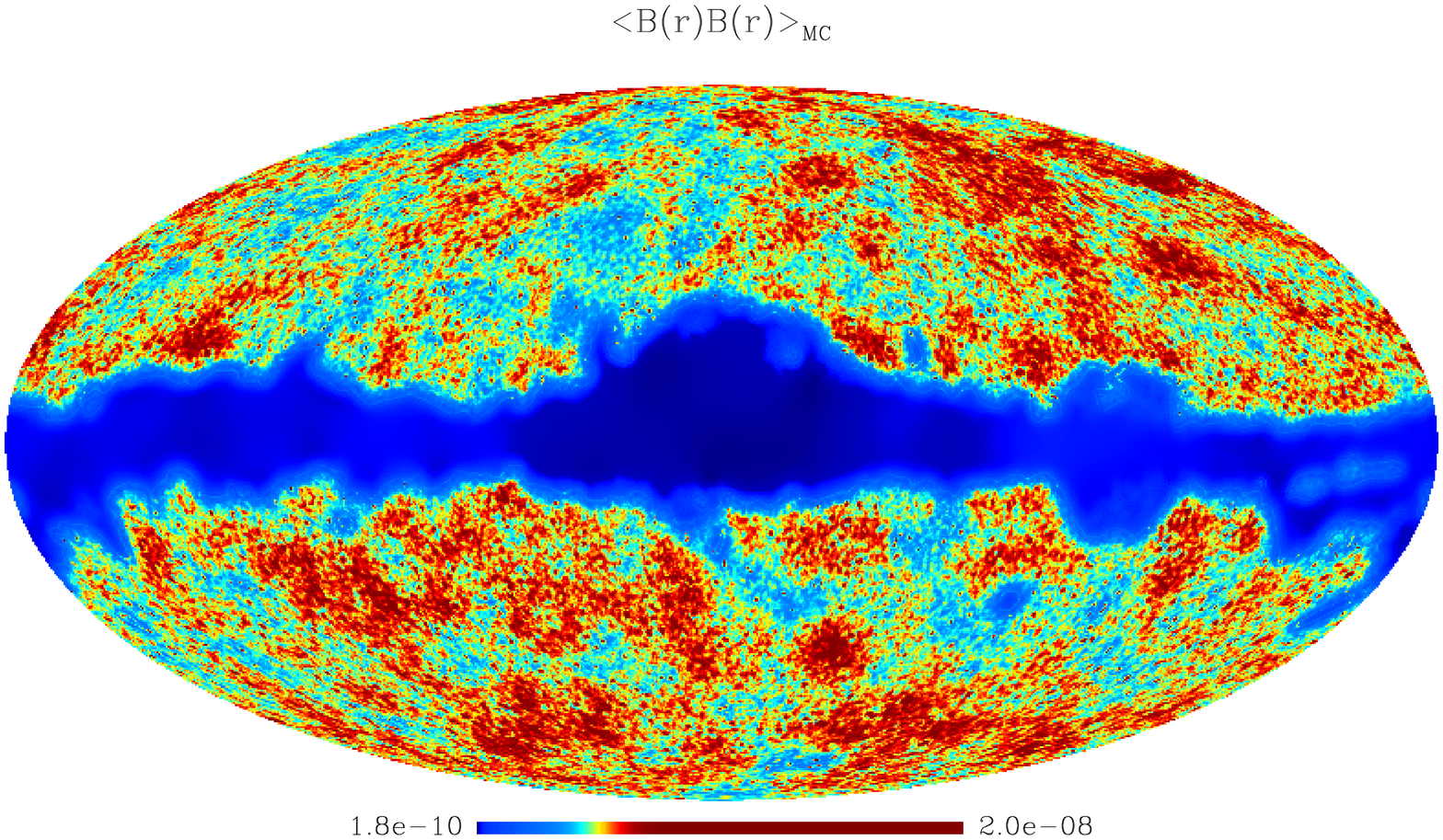}
\caption{The $\langle A_{sim}(\hat{n},r)B_{sim}(\hat{n},r)\rangle_{MC}$
 and  $\langle B^2_{sim}(\hat{n},r)\rangle_{MC}$ maps in dimension-less
 units for a slice near the  surface of  last scattering. These maps are
 calculated from Monte Carlo simulations with the Gaussian signal,
 Planck inhomogeneous noise, and WMAP Kp0 and P06 masks.} 
\label{sabb}
\end{center}
\end{figure*}

Figures~\ref{sabb} and~\ref{noise_sabb} show the maps $\langle
A_{sim}(\hat{n},r)B_{sim}(\hat{n},r)\rangle_{MC} $ and $\langle
B_{sim}^2(\hat{n},r)\rangle_{MC}$, which appear in the linear term
(Eq.~\ref{slin}) of the estimator. These maps are calculated using 100
Monte Carlo simulations of the data. Since the linear term contributes
only in the presence of inhomogeneities, we also show these maps
calculated with noise-only simulations (i.e. no signal). Notice how
these maps correlate with the inhomogeneous noise (as shown in the lower
map of figure~\ref{noise_sabb}).

\section{Conclusion and discussion}
Upcoming CMB experiments will provide a wealth of information about the
CMB polarization anisotropies together with temperature
anisotropies. The combined information from the CMB temperature and
polarization data improves the sensitivity to primordial
non-Gaussianity~\citep{BZ04,YW05,YKW07}. The promise of learning about
the early universe by constraining the amplitude of primordial non-Gaussianity is now well established. In this paper we have generalized the bispectrum estimator of non-Gaussianity described in~\cite{YKW07}, to deal with the inhomogeneous nature of noise and incomplete sky coverage.

The generalization from \cite{YKW07} enables us to increase optimality
of the estimator significantly, without compromising the computational
efficiency of the estimator: the estimator is still computationally
efficient, scaling as $O(N_{pix}^{3/2})$ compared to the
$O(N_{pix}^{5/2})$ scaling of the full bispectrum~\citep{BZ04}
calculation for sky maps with $N_{pix}$ pixels. For the Planck
satellite, this translates into a speed-up by factors of millions,
reducing the required computing time from thousands of years to just
hours and thus making $f_{NL}$ estimation feasible. The speed of our
estimator allows us to study its statistical properties using Monte
Carlo simulations. 

We have used Gaussian and non-Gaussian simulations to
characterize the estimator. We have shown that the generalized fast
estimator is able to deal with the partial sky coverage very well and in
fact the variance of $f_{NL}$ saturates the Fisher matrix bound. In the
presence of both the realistic noise and galactic mask, we find that the
generalized estimator greatly reduces the variance in comparison to
the~\cite{YKW07} estimator of non-Gaussianity using combined CMB
temperature and polarization data. 

Since the estimator is able to deal with the partial sky coverage very effectively, the estimator can also be used to constrain primordial non-Gaussianity using the data from ground and balloon based CMB experiments which observe only a small fraction of the sky. The estimator also solves the problem~\citep{YKW07} of non-trivial polarization mode coupling due to foreground masks. Earlier this issue was dealt with by removing the most contaminated $\ell$ modes from the analysis (usually $\ell <30 $).         

The naive approach of using galactic masks to deal with the polarization contamination is to be refined. Both temperature and polarization foregrounds are expected to produce non-Gaussian signals. Some sources of non-primordial non-Gaussianity are CMB lensing, point sources, and the Sunyaev Zel'dovich effect. Understanding the non-Gaussianity from the polarization foreground sources and refining the estimator to be able to deal with it will be the subject of our future work.

\acknowledgments Some of the  results in this
paper  have  been derived  using  the CMBFAST package by Uros
Seljak and Matias  Zaldarriaga \citep{5} and the  HEALPix package \citep{Healpix}. This work was partially supported by the National Center for Supercomputing Applications under TG-MCA04T015 and by University of Illinois. We also utilized the Teragrid Cluster (www.teragrid.org) at NCSA. BDW acknowledges the Friedrich Wilhelm Bessel research award by the Alexander von Humboldt foundation. BDW and APSY also thank the Max Planck Institute for Astrophysics
for hospitality. BDW and APSY are supported in part by NSF grant numbers
AST 0507676 and 0708849, NASA/JPL subcontract no. 1236748. EK
acknowledges support from the Alfred P. Sloan Foundation.

\bibliographystyle{apj}

\end{document}